\documentstyle[aps,eqsecnum]{revtex}

\begin{document}
\tighten

\wideabs{
\date{July 12, 1999} 
\title{Gravitational waves and the cosmological equation of state}
\author{Teviet Creighton}
\address{Theoretical Astrophysics 130-33, California Institute of
Technology, Pasadena, CA 91125}
\maketitle

\widetext
\begin{abstract}
\hfil
\parbox{5.6in}{As is well known, primordial gravitational waves may be
amplified to detectable levels by parametric amplification during eras
when their wavelengths are pushed outside the cosmological horizon;
this can occur in both inflationary and ``pre-big-bang'' or ``bounce''
cosmologies.  The spectrum of a gravitational wave background is
expressed as a normalized spectral energy density
$\Omega(\omega)\equiv(\omega/\rho_c)(d\rho_{\mathrm{gw}}/d\omega)$,
where $\rho_c$ is the critical energy density that makes the universe
spatially flat, and $d\rho_{\mathrm{gw}}$ is the energy density of
gravitational waves in a frequency band $d\omega$.  The logarithmic
slope of $\Omega$ is simply related to three properties of the early
universe: (i) the gravitons' mean initial quantum occupation number
$N(n)$ ($\equiv1/2$ for a vacuum state), where $n=a\omega$ is the
(invariant) conformal frequency of the mode and $a$ is the
cosmological scale factor, and (ii) \& (iii) the parameter
$\gamma\equiv p/\rho$ of the cosmological equation of state during the
epoch when the waves left the horizon ($\gamma=\gamma_i$) and when
they reentered ($\gamma=\gamma_f$).  In the case of an inflationary
cosmology, the spectral index is
$$
\frac{d\ln\Omega}{d\ln\omega} = \frac{d\ln N}{d\ln n} +
        2\left(\frac{\gamma_i + 1}{\gamma_i + 1/3}\right) +
        2\left(\frac{\gamma_f - 1/3}{\gamma_f + 1/3}\right) \; ,
$$
and for bounce cosmologies it is
$$
\frac{d\ln\Omega}{d\ln\omega} = \frac{d\ln N}{d\ln n} +
        2\left(\frac{2\gamma_i}{\gamma_i + 1/3}\right) +
        2\left(\frac{\gamma_f - 1/3}{\gamma_f + 1/3}\right) \; .
$$
These expressions are compared against various more model-specific
results given in the literature.}
\end{abstract}
\pacs{}}

\narrowtext

\section{Introduction}

The prospect of a detectable cosmological background of gravitational
waves has opened up a new avenue for the investigation of fundamental
physics and cosmology.  Gravitational waves couple very weakly to
matter, and are not blocked or thermalized during any cosmological
epoch back to Planckian densities.  The waves \emph{do}, however,
couple to large-scale cosmological spacetime curvature, and can
therefore provide information about the evolution of this curvature,
and hence about the fundamental physics of the matter fields driving
this evolution.

A great deal of work has been done on this subject, deriving
gravitational-wave spectra from various cosmological models, usually
in hope of producing a model that would generate a detectable
signature in planned gravitational-wave detectors.  The most
comprehensive of these is a recent paper by
Gasperini~\cite{Gasperini_M:1996}, which presents a general
prescription for constructing gravitational-wave spectra from fairly
arbitrary cosmologies.  The results in the present paper are
consistent with Gasperini's results, although I derive them using a
different formalism and consider a more generic cosmology.

The primary intent of this paper, however, is more the reverse of
this: rather than deriving a spectrum from a particular cosmological
model, this paper shows how generic properties of the early universe
can be immediately deduced from an observation of a gravitational-wave
background in any spectral band.  To this end, I have chosen as my
observable the spectral characteristic that is least susceptible to
the tunings of particular cosmological models --- namely, the spectral
index --- and have expressed it in terms of quantities that are most
directly related to the underlying physics of the cosmological
``fluid'' --- namely, the equations of state $p=\gamma\rho$.
Furthermore, since the initial state of the gravitational waves
(before amplification) is potentially one of the more interesting
characteristics that might be deduced from observations, I have left
it as a free parameter, rather than making the usual assumption of
starting the waves in a quantum ground state.

\subsection{Organization of this paper}

In Sec.~\ref{s:derivation}, I present a skeleton of the derivation of
the main formulae of this paper.  Much of the derivation has been done
in one form or another in the published literature, and those familiar
with the field will find nothing surprising, except perhaps the
consideration of non-vacuum initial conditions in
Sec.~\ref{ss:energy-spectra}.  Since the result is largely a
generalization of the more model-specific formulae found in the
literature, Sec.~\ref{s:comparison} compares this paper's formulae
with those published previously.  Sec.~\ref{s:conclusions} presents
some concluding remarks.

\section{The cosmological and wave equations}
\label{s:derivation}

I consider gravitational waves that are linear perturbations on the
spatially flat FRW metric:
\begin{equation}
\label{eq:frw-metric}
ds^2 = a^2(\eta)[-d\eta^2 + dx^2 + dy^2 + dz^2]
        + h_{\alpha\beta}dx^{\alpha}dx^{\beta} \; ,
\end{equation}
where the conformal time coordinate $\eta$ is related to proper time
$t$ by $dt=a(\eta)d\eta$ (I have chosen units in which $c=1$).  The
cosmology has a perfect fluid source that obeys the instantaneous
equation of state $p=\gamma\rho$, where $\gamma$ need not be constant
but typically evolves slowly ($\gamma'/\gamma\ll a'/a$, where $'\equiv
d/d\eta$).  Causality considerations require that $\gamma\leq1$;
realistically, we can assume $\gamma<1$, since not all of the energy
in the universe will be in the maximally-stiff field.  One can solve
Einstein's equations to zeroth order in $h$ to obtain the exact
solution
\begin{equation}
\label{eq:frw-evolution}
a(\eta) = a_0 \exp\left\{\int_{\eta_0}^\eta \frac{d\eta_1}
	{a_0/a'_0 + \int_{\eta_0}^{\eta_1} d\eta_2
		[1+3\gamma(\eta_2)]/2} \right\} \; ,
\end{equation}
which reduces to the usual power-law evolution
$a(\eta)\sim\eta^{2/(1+3\gamma)}$ during epochs of (nearly) constant
$\gamma$.

Following the notation in Eq.~(2) of~\cite{Grishchuk_LP:1991}, I write
the linear perturbations in the form:
\begin{equation}
\label{eq:h-def}
h_{\alpha\beta} = \sum_{{\mathbf n},j} \frac{\mu_n(\eta)}{a(\eta)}
	U_{\mathbf n}({\mathbf x}) e^{(j)}_{\alpha\beta} \; ,
\end{equation}
where $e^{(j)}_{\alpha\beta}$ is some basis of polarization tensors,
$U_{\mathbf n}({\mathbf x})\propto e^{i{\mathbf n}\cdot{\mathbf x}}$
is a spatial harmonic function with a true (physical) wave number
${\mathbf k}={\mathbf n}/a$ and frequency $\omega=n/a$
($n=\|\mathbf{n}\|$), and $\mu_n(\eta)$ obeys the Schr\"odinger-like
equation:
\begin{equation}
\label{eq:wave-eqn}
\mu_n'' + \left(n^2 - \frac{a''}{a}\right)\mu_n = 0 \; .
\end{equation}
Physically, the relative magnitudes of $n^2$ and the effective
potential $a''/a$ are related to whether the wave is inside or outside
of the Hubble radius.  Comparing $\mathbf k$ to the Hubble radius
$r_H=a/(da/dt)$, one has:
\begin{equation}
\label{eq:horizon-crossing}
k^2r_H^2 = \frac{n^2}{(a'/a)^2}
	= \frac{n^2}{a''/a}\times\frac{1-3\gamma}{2} \; .
\end{equation}
Roughly speaking, a wave that is hitting the effective potential
($a''/a$ increasing to meet $n^2$) is one that is exiting the Hubble
radius, to order of magnitude.  Similarly, a wave that is emerging
from the effective potiential is reentering the Hubble radius.  The
exception is for epochs when $\gamma\rightarrow1/3$; see the remarks
near the end of Sec.~\ref{ss:n-dependence}.

\begin{figure}
\setlength{\unitlength}{2cm}
\begin{picture}(4,1.3)
\put(3,0.3){

\qbezier[50](-3,0.09)(-1.5,0.18)(-1,0.8)
\qbezier[50](-1,0.8)(-0.25,0.26)(1,0.2)
\put(-0.6,0.65){$(a'/a)^2$}

\qbezier(-3,0.11)(-1.5,0.22)(-1,1)
\qbezier(-1,1)(-1,0.75)(-1,0.7)
\qbezier(-1,0.7)(-0.25,0.23)(1,0.18)
\put(-1.4,0.65){\hspace{-3em}$|a''/a|$}

\multiput(-3,0.35)(0.1,0){40}{\line(1,0){0.05}}
\put(1,0.35){\hspace{-1em}\raisebox{0.5ex}{$n^2$}}

\put(-3,0){\line(1,0){4}}

\put(-1.85,0){\line(0,1){0.03}}
\put(-1.85,0){\raisebox{-2ex}{\hspace{-0.25em}$\eta_i$}}
\put(-1,0){\line(0,1){0.03}}
\put(-1,0){\raisebox{-2ex}{\hspace{-0.25em}$\eta_1$}}
\put(-0.2,0){\line(0,1){0.03}}
\put(-0.2,0){\raisebox{-2ex}{\hspace{-0.25em}$\eta_f$}}
}
\end{picture}
\caption{\label{fig:simple-barrier} Schematic of the evolution of the
effective potential $|a''/a|$ (sollid line) and Hubble scale
$(a'/a)^2$ (dotted line) for inflation ($\eta<\eta_1$) followed by
decelerating expansion ($\eta>\eta_1$).  Also shown are the times
$\eta_i$ and $\eta_f$ when a wave with squared conformal frequency
$n^2$ (dashed line) hits and emerges from the potential barrier.}
\end{figure}
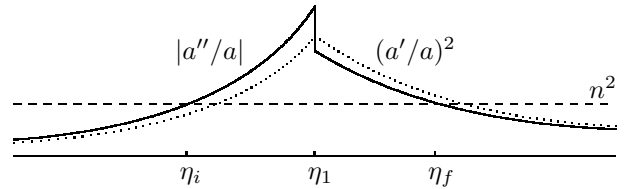

Cosmological amplification of waves occurs where initially oscillatory
waves hit the effective potential during epochs of accelerated
collapse ($a'<0$, $\gamma>-1/3$) or accelerated expansion ($a'>0$,
$\gamma<-1/3$), when $|a''/a|$ is increasing, and then emerge into a
post-inflationary universe ($a'>0$, $\gamma>-1/3$) in which $|a''/a|$
is decreasing.  A schematic of such an evolution is shown in
Fig.~\ref{fig:simple-barrier}.  The periods when the wave initially
hits the potential and when it finally emerges from it will be denoted
by subscript $i$ and $f$, respectively.  The (complex) amplification
factor of the emerging waves
$\beta_n\equiv\mu_n(\eta_f)/\mu_n(\eta_i)$ is
\begin{eqnarray}
\beta_n & = & \frac{1}{2i}e^{in(\eta_i-\eta_f)}\left[
	\frac{a_f}{a_i}\left(i+\frac{a'_f}{a_f n}\right) +
	\frac{a_i}{a_f}\left(i-\frac{a'_i}{a_i n}\right)
	\right. \nonumber\\
\label{eq:beta} & & \left.
	{} + a_i a_f n \left(i-\frac{a'_i}{a_i n}\right)
		\left(i+\frac{a'_f}{a_f n}\right)
		\int_{\eta_i}^{\eta_f} \frac{d\eta}{a^2}\right] \; ;
\end{eqnarray}
this is Eq.~(11) of~\cite{Grishchuk_LP:1991} with a sign error
corrected, and reexpressed in the current notation.

If the evolution of $a(\eta)$ between $a_i$ and $a_f$ is monotonic,
one can follow~\cite{Grishchuk_LP:1991} and use the equation of state
to reexpress the integral over $\eta$ as an integral over $a$,
eventually obtaining, for $a_f\gg a_i$:
\begin{equation}
\label{eq:beta-monotonic}
|\beta_n| \approx \frac{1}{2}\frac{a_f}{a_i}\left|
        \left(i+\frac{a'_f}{a_f n}\right)
        \left(1 - \frac{2/3}{1-\langle\gamma\rangle}
        \left[1 - \frac{i a_i n}{a'_i}\right]\right)\right| \; ,
\end{equation}
where $\langle\gamma\rangle$ is the average of $\gamma$ from $\eta_i$
to $\eta_f$ with weighting factor $1/a^2$.

The derivation of Eq.~(\ref{eq:beta-monotonic}) breaks down for the
case of a bounce cosmology, where $a(\eta)$ is not monotonic.  In this
case, however, Eq.~(\ref{eq:beta}) is almost entirely dominated by the
contribution of the integral $\int d\eta/a^2$ near the time of the
bounce.  Modeling the bounce as smooth with some finite concavity
$a''_0$ around the minimum $a_0$, we obtain:
\begin{equation}
\label{eq:beta-bounce}
|\beta_n| \approx \frac{1}{2} a_i a_f n \left|
        \left(i+\frac{a'_f}{a_f n}\right)
        \left(i-\frac{a'_i}{a_i n}\right)\right|
        \frac{1}{\sqrt{a''_0a_0^3}} \; .
\end{equation}

\subsection{Dependence on $n$}
\label{ss:n-dependence}

The amplification factor $\beta_n$ is at least implicitly dependent on
$n$, since waves of different $n$ will hit and emerge from the
effective potential at different times $\eta_i$ and $\eta_f$, and
hence with different $a_i$, $a_f$.  For periods of constant $\gamma$,
we have $a\sim\eta^{2/(1+3\gamma)}$.  The condition $n^2=a''/a$ for
hitting or emerging from the effective potential implies that
$a_{i,f}\sim n^{-2/(1+3\gamma_{i,f})}$, and that
$a'_{i,f}/(a_{i,f}n)\sim\sqrt{2/|1-3\gamma_{i,f}|}\sim\mathrm{constant}$.
If $\gamma$ is changing gradually near $\eta_{i,f}$ these formulae
remain true up to corrections of order $(\gamma'/\gamma)(a'/a)^{-1}$.
Furthermore, the weighted average $\langle\gamma\rangle$ in
Eq.~(\ref{eq:beta-monotonic}) depends only very weakly on
$\eta_{i,f}$, and hence on $n$, except for values of $n^2$ near the
peak of the effective potential.  These considerations, combined with
Eqs.~(\ref{eq:beta-monotonic}) and~(\ref{eq:beta-bounce}), give
\begin{equation}
\label{eq:beta-n-monotonic}
|\beta_n| \sim n^{2/(1+3\gamma_i)-2/(1+3\gamma_f)}
\end{equation}
for inflationary cosmologies, and
\begin{equation}
\label{eq:beta-n-bounce}
|\beta_n| \sim n^{1+2/(1+3\gamma_i)+2/(1+3\gamma_f)}
\end{equation}
for bounce cosmologies.

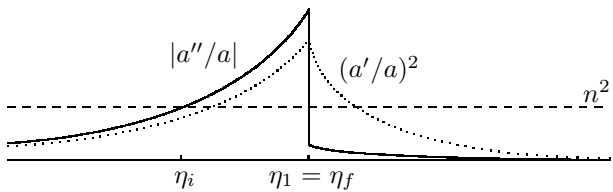
\begin{figure}
\setlength{\unitlength}{2cm}
\begin{picture}(4,1.3)
\put(3,0.3){

\qbezier[50](-3,0.09)(-1.5,0.18)(-1,0.8)
\qbezier[50](-1,0.8)(-0.86,0.03)(1,0.01)
\put(-0.8,0.55){$(a'/a)^2$}

\qbezier(-3,0.11)(-1.5,0.22)(-1,1)
\qbezier(-1,1)(-1,0.5)(-1,0.1)
\qbezier(-1,0.1)(-0.86,0.01)(1,0)
\put(-1.4,0.65){\hspace{-3em}$|a''/a|$}

\multiput(-3,0.35)(0.1,0){40}{\line(1,0){0.05}}
\put(1,0.35){\hspace{-1em}\raisebox{0.5ex}{$n^2$}}

\put(-3,0){\line(1,0){4}}

\put(-1.85,0){\line(0,1){0.03}}
\put(-1.85,0){\raisebox{-2ex}{\hspace{-0.25em}$\eta_i$}}
\put(-1,0){\line(0,1){0.03}}
\put(-1,0){\raisebox{-2ex}{\hspace{-1.5em}$\eta_1=\eta_f$}}
}
\end{picture}
\caption{\label{fig:radiation-dominated} Schematic of the evolution of
the effective potential and Hubble scale for inflation ($\eta<\eta_1$)
followed by a radiation-dominated equation of state ($\eta>\eta_1$).
Note that $\eta_f=\eta_1$ independent of $n$ for a broad range of
frequencies.}
\end{figure}

A special case is when $\gamma_f\approx1/3$ (the equation of state for
a relativistic or radiation-dominated fluid), for which the effective
potential $a''/a\approx0$.  This is depicted in
Fig.~\ref{fig:radiation-dominated}.  If the transition to radiation
dominance is rapid, then $\eta_f$, and hence $a_f$ and $a'_f$, are
independent of $n$ for a wide range of $n$.  However, waves that are
well outside the Hubble radius at the moment of transition will have
$a'_f/a_f\gg n$, as is clear from Eq.~(\ref{eq:horizon-crossing}) and
Fig.~\ref{fig:radiation-dominated}.  So the appropriate terms in
Eqs.~(\ref{eq:beta-monotonic}) and~(\ref{eq:beta-bounce}) have
behaviour
\begin{equation}
\label{eq:radiation-dependence}
a_f\left|i+\frac{a'_f}{a_f n}\right| \sim \frac{1}{n} \; .
\end{equation}
This is the same dependence as one would get by na\"{\i}vely plugging
$\gamma_f=1/3$ into Eqs.~(\ref{eq:beta-n-monotonic})
and~(\ref{eq:beta-n-bounce}), so these equations are correct even as
$\gamma_f\rightarrow1/3$.  (This argument also applies for waves that
\emph{hit} the effective potential during radiation-dominated
collapse.)

\begin{figure}
\setlength{\unitlength}{2cm}
\begin{picture}(4,1.3)
\put(3,0.3){

\qbezier[50](-3,0.09)(-1.5,0.18)(-1,0.8)
\qbezier[10](-1,0.8)(-0.9,0.6)(-0.8,0.55)
\qbezier[40](-0.8,0.55)(0.173,0.31)(1,0.24)
\put(-0.8,0.65){$(a'/a)^2$}

\qbezier(-3,0.11)(-1.5,0.22)(-1,1)
\qbezier(-1,1)(-1,0.5)(-1,0.1)
\qbezier(-1,0.1)(-0.9,0.06)(-0.8,0.05)
\qbezier(-0.8,0.05)(-0.8,0.1)(-0.8,0.48)
\qbezier(-0.8,0.48)(0.173,0.27)(1,0.2)
\put(-1.4,0.65){\hspace{-3em}$|a''/a|$}

\multiput(-3,0.35)(0.1,0){40}{\line(1,0){0.05}}
\put(1,0.35){\hspace{-1em}\raisebox{0.5ex}{$n^2$}}

\put(-3,0){\line(1,0){4}}

\put(-1.85,0){\line(0,1){0.03}}
\put(-1.85,0){\raisebox{-2ex}{\hspace{-0.25em}$\eta_i$}}
\put(-1,0){\line(0,1){0.03}}
\put(-1,0){\raisebox{-2ex}{\hspace{-0.25em}$\eta_1$}}
\put(-0.8,0){\line(0,1){0.03}}
\put(-0.8,0){\raisebox{-2ex}{\hspace{-0.25em}$\eta_2$}}
\put(-0.1,0){\line(0,1){0.03}}
\put(-0.1,0){\raisebox{-2ex}{\hspace{-0.25em}$\eta_f$}}
}
\end{picture}
\caption{\label{fig:three-phase} Schematic of the evolution of the
effective potential and Hubble scale for inflation ($\eta<\eta_1$),
followed by radiation dominance ($\eta_1<\eta<\eta_2$), followed by
generic decelerated expansion ($\eta>\eta_2$).  Waves emerging from
the potential barrier after $\eta_2$ will have been outside the Hubble
radius throughout the period of radiation dominance.}
\end{figure}
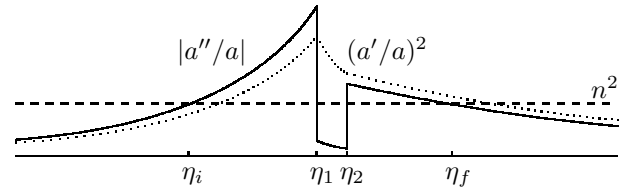

A similar situation occurs for waves that remain outside the Hubble
radius throughout an intermediate period of $\gamma\approx1/3$, when
formally they have emerged from the effective potential and are
oscillatory.  This is depicted in Fig.~\ref{fig:three-phase}.  The
effect of an intermediate period $\Delta\eta$ of zero effective
potential is a term $\sim\sin(n\Delta\eta)/n$ in the final amplitude.
If the waves remain well outside the Hubble radius throughout this
era, then $n\Delta\eta\ll1$, and this term is a constant with respect
to $n$.  Once again, Eqs.~(\ref{eq:beta-n-monotonic})
and~(\ref{eq:beta-n-bounce}) remain unchanged.  Both this and the
preceding cases confirm one's physical intuition, that the final
spectrum cannot resolve the details of sudden phase transitions that
occur when the waves are much larger than the Hubble radius.

Eqs.~(\ref{eq:beta-n-monotonic}), (\ref{eq:beta-n-bounce}) also break
down when $\gamma_{i,f}\approx-1/3$, but this condition will never
arise.  When $\gamma\approx-1/3$, one has $a\sim e^{\kappa\eta}$, so
$a''/a\approx\mathrm{constant}$, and waves will neither hit nor leave
the effective potential.

\subsection{The initial and final spectra}
\label{ss:energy-spectra}

Most analyses of primordial gravitational waves assume that the
initial state of the metric perturbations (during the Planck era or
during pre-Big-Bang collapse) is a quantum-mechanical vacuum.  While
this is a reasonable assumption, one does not \emph{know} the initial
state in advance of observation.  In fact, primordial gravitational
waves are a potential means of observing the initial state of the
cosmos.  It is instructive, therefore, to leave the initial graviton
spectrum as a free parameter.

I parameterize the initial spectrum using the mean quantum occupation
number of graviton modes as a function of mode frequency, $N(n)$.  In
the semiclassical approach, the creation of gravitons can be treated
as the classical amplification of the vacuum energy $\hbar\omega/2$ in
each mode, so I normalize $N(n)$ to be $1/2$ for a mode in the ground
state (no real gravitons), $3/2$ for the first excited state (one real
graviton), and so on.  This approach is valid provided the final state
is classical ($|\beta_n|\gg1$).

Cosmological perturbation spectra are normally expressed as a
normalized energy density per logarithmic frequency interval,
$\Omega\equiv(dE/dV d\ln\omega)/\rho_c$ (where $\rho_c$ is the energy
density required to make the universe spatially flat).  There are
$2\omega^2 d\omega dV$ modes in a frequency interval $d\omega$ in a
volume $dV$, each with energy $\hbar\omega N(n)$ (where $n=a\omega$),
so the initial energy spectrum is:
\begin{equation}
\label{eq:initial-energy}
\Omega_{\mathrm initial}(n) = \frac{2n^4 N(n)}
	{a_{\mathrm initial}^4\rho_c} \; .
\end{equation}
Note that $a_{\mathrm initial}$ is the (constant) scale factor of the
universe at the time that one is evaluating the initial spectrum; it
is not the same as $a_i$, which depends on $n$.

In considering the logarithmic slope of the spectrum I ignore overall
amplitude factors, which are highly model-specific.  The energy
spectrum at the present day is quadratic in $|\beta_n|$, so its
logarithmic slope is:
\begin{equation}
\label{eq:final-energy}
\frac{d\ln\Omega}{d\ln\omega} =
	\frac{d\ln\Omega_{\mathrm{initial}}}{d\ln n} +
        2\frac{d\ln |\beta_n|}{d\ln n} \; ,
\end{equation}
where the evaluation is at $n=\omega/a_{\mathrm present}$.  Combining
this with Eqs.~(\ref{eq:initial-energy}), (\ref{eq:beta-n-monotonic}),
and~(\ref{eq:beta-n-bounce}) gives the pricipal result of this paper:
\begin{equation}
\label{eq:spectrum-monotonic}
\frac{d\ln\Omega}{d\ln\omega} = \frac{d\ln N}{d\ln n} +
        2\left(\frac{\gamma_i + 1}{\gamma_i + 1/3}\right) +
        2\left(\frac{\gamma_f - 1/3}{\gamma_f + 1/3}\right) 
\end{equation}
for an inflationary cosmology, and:
\begin{equation}
\label{eq:spectrum-bounce}
\frac{d\ln\Omega}{d\ln\omega} = \frac{d\ln N}{d\ln n} +
        2\left(\frac{2\gamma_i}{\gamma_i + 1/3}\right) +
        2\left(\frac{\gamma_f - 1/3}{\gamma_f + 1/3}\right)
\end{equation}
for a bounce-type cosmology.

\section{Comparison with special cases}
\label{s:comparison}

The formulae given above are consistent with results established
previously in the literature.  I first show this for the two
``standard'' cases of deSitter inflation and of cosmological rebound
from a state of weakly coupled strings.  I then compare with recent
papers that consider more generic pre- and post-inflationary
evolutions.

\subsection{Standard inflation}
\label{ss:standard-inflation}

The ``standard'' model for an inflationary cosmology consists of a
period of deSitter expansion ($\gamma=-1$) due to the universe being
in a false vacuum with nonzero energy density.  In the
post-inflationary era, the universe quickly reheats to a
radiation-dominated equation of state ($\gamma=1/3$), followed by
cooling to a matter-dominated equation of state ($\gamma=0$).  If we
assume the initial graviton state from the Planck era was a vacuum,
then for waves that emerged into the radiation-dominated universe we
have $N\equiv1/2$, $\gamma_i=-1$, $\gamma_f=1/3$, and
\begin{equation}
\label{eq:standard-inflation}
\frac{d\ln\Omega}{d\ln\omega} = 0
\end{equation}
This is, of course, the flat spectrum predicted by Harrison and
Zel'dovich~\cite{Harrison_ER:1970,Zeldovich_YB:1972}.

The last two decades of the spectrum ($10^{-16}$--$10^{-18}$~Hz)
consist of waves that emerged more recently, when the universe was
matter-dominated ($\gamma_f=0$).  In this band the spectrum is tilted
towards low frequencies, with a spectral index of $-2$.

\subsection{Standard string-motivated rebound cosmologies}
\label{ss:standard-string}

It has been shown~\cite{Brustein_R:1995,Gasperini_M:1995} that string
theories can lead to a gravitational-wave background that is
increasing with frequency.  These models are consistent with the
current derivation provided one recognizes that the current formalism
applies to the Einstein frame (where the metric represents intervals
measured by physical rulers and clocks), rather than the string frame
(where the metric geodesics describe the trajectories of weakly
coupled strings).  In the Einstein frame, a typical evolution consists
of an epoch of decreasing lengthscale $a$ and weakly coupled strings,
followed by a period of strongly coupled strings that reverse the
collapse, followed by relaxation of the string dilaton and an
expanding radiation-dominated universe, eventually cooling to matter
dominance.  As usual, a vacuum initial state ($N\equiv1/2$) is
normally assumed.  Most modes hit the effective potential during the
epoch of weakly interacting strings, when the kinetic term of the
dilaton dominates the energy background, giving a nearly maximally
stiff equation of state ($\gamma_i\approx1$).  For waves that emerge
during the radiation-dominated phase ($\gamma_f\approx1/3$),
Eq.~(\ref{eq:spectrum-bounce}) gives:
\begin{equation}
\label{eq:standard-string}
\frac{d\ln\Omega}{d\ln\omega} = 3 \; .
\end{equation}
Eq.~(3.3) of~\cite{Brustein_R:1995} and Eq.~(5.7)
of~\cite{Gasperini_M:1995} also give a spectral index of 3 for low
frequencies.  These papers also predict a logarithmic cutoff at high
frequencies, $\Omega\sim\omega^3\ln^2(\omega_s/\omega)$, where
$\omega_s$ is the maximum frequency of waves that encountered the
effective potential during the epoch of maximal stiffness (typically
of order $10^{11}$~Hz, depending on the details of the strongly
coupled string epoch).  This effect occurs only for $\gamma_i$ exactly
equal to 1, which the current analysis doesn't consider, and only
affects the spectral index within a decade of $\omega_s$ in any case.

The low-frequency tail ($\omega<10^{-16}$~Hz) of waves that emerge
when $\gamma_f=0$ will also be tilted towards high frequencies, but
with a spectral index of 1.

\subsection{Nonstandard equations of state}
\label{ss:stiff-pre-radiation}

A recent paper~\cite{Giovannini_M:1998} has explored the effects of an
intermediate phase between the inflationary and radiation-dominated
eras, in which the dominant cosmological fluid had an equation of
state stiffer than radiation ($1/3<\gamma_f<1$).  One consequence is
that the background gravitational-wave spectrum would be tilted
towards high frequencies for those wavelengths that emerged during
this era.  Assuming an initial vacuum ($N\equiv1/2$) and deSitter
inflation, Eq.~(\ref{eq:spectrum-monotonic}) gives us:
\begin{equation}
\label{eq:stiff-pre-radiation}
\frac{d\ln\Omega}{d\ln\omega} = 2\frac{\gamma_f - 1/3}{\gamma_f + 1/3}
        \; .
\end{equation}
Eq.~(3.32) of~\cite{Giovannini_M:1998} gives the same dependence on
$\gamma$.

Eq.~(3.31) of~\cite{Giovannini_M:1998} extends the analysis to the
case of a maximally-stiff equation of state ($\gamma_f=1$), which this
paper does not consider.  This gives a high-frequency logarithmic
cutoff $\Omega\sim\omega\ln^2(\omega_1/\omega)$, where $\omega_1$ is
the highest frequency that encountered the amplifying potential.  The
cutoff does not affect the spectral index at frequencies more than a
decade or so below $\omega_1$, which is likely in the MHz to GHz range
(well above the pass-bands of proposed gravitational-wave detectors).
This is analogous to the situation mentioned earlier, when
$\gamma_i\rightarrow1$ during collapse.

Gasperini~\cite{Gasperini_M:1996} has performed an even more general
analysis of the primordial gravitational wave spectrum, considering
generic evolution of $a(\eta)$ both in the early and late cosmological
epochs.  As usual, the initial state of the waves was taken to be a
vacuum ($N\equiv1/2$).  Eqs.~(\ref{eq:spectrum-monotonic}),
(\ref{eq:spectrum-bounce}) can then be written as:
\begin{equation}
\label{eq:vacuum-monotonic}
\frac{d\ln\Omega}{d\ln\omega} =
        4+\frac{4}{1+3\gamma_i}-\frac{4}{1+3\gamma_f}
\end{equation}
or
\begin{equation}
\label{eq:vacuum-bounce}
\frac{d\ln\Omega}{d\ln\omega} =
        6-\frac{4}{1+3\gamma_i}-\frac{4}{1+3\gamma_f}
\end{equation}
for monotonic or bounce cosmologies, respectively.  By comparison,
Eq.~(4.26) of~\cite{Gasperini_M:1996} gives the spectral index as
$4-2\nu_1-2\nu_{i+1}$, where $\nu=|\alpha-1/2|$ and $\alpha$ is the
exponent of the power law $a\sim\eta^{\alpha}$ during the epoch when
the waves first strike the effective potential (for $\nu_{i+1}$) and
when they leave the effective potential (for $\nu_1$).  Gasperini's
results are identical to Eqs.~(\ref{eq:vacuum-monotonic}) and
(\ref{eq:vacuum-bounce}), provided one recognizes two things.

First, a phase of slowly varying $\alpha$ corresponds to a phase of
slowly varying $\gamma$, with $\alpha=2/(1+3\gamma)$.  Inflationary
cosmologies have $\gamma_i<-1/3$, hence $\alpha_{i+1}<0$.  Bounce
cosmologies have $-1/3<\gamma_i\leq1$ and $\alpha_{i+1}\geq1/2$.
Either cosmology has $-1/3<\gamma_f\leq1$ and $\alpha_1\geq1/2$.  (The
condition $0<\alpha<1/2$ corresponds to the physically unrealistic
case of $\gamma>1$.)

Second, Gasperini assumes in Eq.~(4.26) of~\cite{Gasperini_M:1996}
that all waves leave the effective potential at the same cosmological
epoch, so that $\nu_1$ is a constant across all frequencies (normally
$\nu_1=1/2$ for radiation dominance).  In fact, the spectral index
(but not necessarily the normalization) given in Eq.~(4.26)
of~\cite{Gasperini_M:1996} is valid for $\nu_1$ varying across
frequencies.

\section{Conclusions}
\label{s:conclusions}

Eqs.~(\ref{eq:spectrum-monotonic}) and~(\ref{eq:spectrum-bounce}) show
how much can be determined about early cosmology from an observation
of the spectral index of primordial gravitational waves.  Although
these formulae are quite simple, they still involve three independent
parameters: $\gamma_i$, $\gamma_f$, and $N(n)$.  To make definite
statements about any one of them, one must make assumptions about the
others.  The usual procedure is to assume that we know $N(n)$ and
$\gamma_f$, leaving $\gamma_i$ to be deduced; however, it is quite
possible that future advances in theory will fix both $\gamma_i$ and
$\gamma_f$, allowing gravitational-wave observations to probe directly
the Planck-scale structure of the universe.

The formulae are also useful in showing which cosmological theories
produce equivalent gravitational signatures.  For instance, a
monotonically inflating cosmology with $\gamma_{i_1}<-1/3$ will yield
the same spectral index as a bounce-type cosmology with
$\gamma_{i_2}=(\gamma_{i_1}+1)/(3\gamma_{i_1}-1)$ in the range
$(-1/3,1/3)$.  Since these are physically valid ranges for $\gamma_i$
in each case, it follows that a monotonic cosmology cannot be
distinguished from a bounce cosmology given only the
gravitational-wave spectral index.

The purpose of this paper was to present the simplest possible
expressions relating a generic cosmological gravitational-wave
signature to the physical processes that produced it, with a minimum
of assumptions about the underlying cosmological model.  I have
deliberately chosen an approach of ``maximum ignorance'', allowing, as
much as possible, for the observations to dictate the cosmological
parameters.  Formulae such as these may prove valuable in an era when
gravitational-wave observations begin to explore a new realm of
physics, possibly revealing things beyond our most informed
predictions.

\section*{Acknowledgements}

This work was supported by NSF grant \mbox{AST-9731698} and NASA grant
\mbox{NAG5-6840}.  I thank Kip Thorne for his help and encouragement.

\end{document}